\documentclass[useAMS,usenatbib]{mn2e}

\usepackage[]{suprtab,graphicx}
\def\tabfactor{93}

\newcommand{\etal}      {et al.}
\newcommand{\eg}        {e.g.~}

\def\degrees{$^\circ$}
\def\amin{$^{'}$}
\def\asec{$^{''}$}

\title[A Search for Interstellar Pyrimidine]
      {A Search for Interstellar Pyrimidine}
\author[Y.-J. Kuan, C.-H. Yan, S.B. Charnley, Z. Kisiel, P. Ehrenfreund,
\& H.-C. Huang]
{Yi-Jehng Kuan,$^{1,2}$\thanks{E-mail: kuan@sgrb2.geos.ntnu.edu.tw (YJK)}
 Chi-Hung Yan,$^{1}$ Steven B. Charnley,$^{3}$ Zbigniew Kisiel,$^{4}$
\newauthor
 Pascale Ehrenfreund$^{5}$ and Hui-Chun Huang$^{1}$\\
$^{1}$Department of Earth Sciences, National Taiwan Normal University,
      Taipei 116, Taiwan, ROC \\
$^{2}$Institute of Astronomy and Astrophysics, Academia Sinica,
      P.O. Box 23-141, Taipei 106, Taiwan, ROC \\
$^{3}$Space Science Division, NASA Ames Research Centre, MS 245-3,
      Moffett Field, CA 94035, USA \\
$^{4}$Institute of Physics, Polish Academy of Sciences,
      Al.Lotnikow 32/46, 02-668 Warsaw, Poland\\
$^{5}$Leiden Observatory, P.O. Box 9513, 2300 RA  Leiden, The Netherlands}

\begin{document}

\date{Accepted 2003 June 30. Received 2003 June 17; in original form 2003 April 14}

\pagerange{\pageref{firstpage}--\pageref{lastpage}} \pubyear{2003}

\maketitle

\label{firstpage}

\begin{abstract}
 We have searched three hot molecular cores for submillimeter emission
 from the nucleic acid building-block pyrimidine. We obtain upper
 limits to the total pyrimidine (beam-averaged) column densities
 towards Sgr B2(N), Orion KL and W51 e1/e2 of $1.7\times 10^{14}\rm
 cm^{-2}$, $2.4\times 10^{14}\rm cm^{-2}$ and $3.4\times 10^{14}\rm
 cm^{-2}$, respectively.  The associated upper limits to the
 pyrimidine fractional abundances lie in the range $(0.3-3)\times
 10^{-10}$.  Implications of this result for interstellar organic
 chemistry, and for the prospects of detecting nitrogen heterocycles in
 general, are briefly discussed.

\end{abstract}


\section{Introduction} \label{intro}

 Molecular clouds contain many organic molecules that are known to be
 important in biochemistry. Astronomical observations, particularly at
 radio wavelengths, allow us to determine the chemical composition and
 characteristics of this molecular inventory (\eg \citealt{dic01};
 \citealt{cha01}).  Following incorporation into
 protostellar disks and comets, these molecules, or their descendants,
 were probably the major source of volatile organic material available
 to the early Earth \citep{chy90}. Studies of molecular cloud
 composition therefore enable us to quantitatively address the issue of
 the connection between interstellar chemistry, the organic composition
 of primitive Solar System material, and the origin, evolution and
 distribution of Life in the Galaxy (\eg \citealt{ehr00,ehr02}).

 Many organics that are known, or strongly suspected, to be present in
 interstellar clouds, are fundamental components of the large organic
 macromolecules that are central to biochemistry. Examples of these are
 sugars and amino acids, the respective building blocks of
 polysaccharides and proteins.  The simplest of these, glycolaldehyde
 and glycine, are both identified in the interstellar medium
 \citep{hol00,kua03}.

 Until recently, definitive detections of interstellar ring compounds
 have been scarce (\eg ethylene oxide, \citealt{dic97}) and previous
 searches for imidazole, cyanoform, pyrrole and pyrimidine were
 unsuccessful \citep{sim73,mye80,irv81}.  Recently we have tentatively
 detected the azaheterocyclic compounds 2H-azirine ($c$-$\rm C_2H_3N$)
 and aziridine ($c$-$\rm C_2H_5N$) at mm wavelengths \citep{kub03,cha01}.
 Two (different) tentative lines of Aziridine have also been claimed by
 \citet{dic01}.  These observations suggest that biochemically
 important ring molecules may await detection. Of these potential
 discoveries, a key interstellar molecule for Astrobiology, comparable
 in importance to glycine, would be pyrimidine ($c$-$\rm C_4H_4N_2$),
 the unsubstituted
 ring analogue for three of the DNA and RNA bases: thymine, cytosine and
 uracil. Interstellar pyrimidine was unsuccessfully searched for at 46 GHz
 30 years ago \citep{sim73}. Evidence for various purines and pyrimidines
 in space, including pyrimidine, comes from the fact that they have been
 detected in meteoritic organic matter \citep{sto81,sto82}, and also may be
 components of Comet Halley's CHON dust \citep{kru91}.

 The molecular composition of hot molecular cores is known to largely
 reflect the solid state chemistry that occurred on grain surfaces,
 prior to their deposition into the gas after protostellar dust heating
 (see \citealt{ehr00}).  Reactions between HCN and VyCN ($\rm CH_2CHCN$)
 on the surface of dust grains has been suggested as a possible source of
 interstellar pyrimidine \citep{sim73}.  Observations of hot cores, known
 to contain other rings believed to have formed on dust prior to
 evaporation, such as $c$-$\rm C_2H_4O$ (ethylene oxide) \citep{dic97,num98}
 and $c$-$\rm C_2H_3N$ \citep{cha01}, do show the very high abundances of
 the HCN and VyCN molecules required for formation of N-heterocycles
 \citep{ziu86,tur91,num99}.  Alternatively, experiments indicate that
 energetic processing of ices (\eg \citealt{all97}) leads to an organic
 residue containing many polycyclic aromatic hydrocarbons (PAHs)
 \citep{gre00}.  One may speculate that similar processing of interstellar
 ice analogues containing ammonia, molecular nitrogen, and hydrogen cyanide,
 may similarly produce N-heterocycles like pyrimidine.

 Hence, searches of hot molecular cores, employing high-quality
 spectroscopic data, may offer the best chance of a pyrimidine
 detection.  In this paper we report the results of a search towards
 three massive star-forming regions for submillimeter emission from
 pyrimidine.

\section{Observations}

 We carried out a program of submillimeter observations using the
 James Clerk Maxwell Telescope (JCMT)\footnote{The JCMT is operated by
 the Joint Astronomy Centre in Hilo, Hawaii on behalf of the present
 organizations: the Particle Physics and Astronomy Research Council in
 the United Kingdom, The National Raserch Coucnil of Canada and the
 Netherlands Organization for Scientific Research} on Mauna Kea,
 Hawaii during Semester 01B over the period September 9--13, 2001.
 Additional measurements were made on November 23, December 1--2, and
 December 10, 2001.  Our target list is given in Table~\ref{tbl-1};
 Column 4 gives the nominal LSR velocity of each source adopted for
 the search.  The source list consists of three well-studied regions
 of massive star formation which are known to be rich in complex
 organic molecules (\eg \citealt{bla87}; \citealt{mia95};
 \citealt{rem02}).

\begin{table}
\caption{Source List}\label{tbl-1}
\begin{tabular}{lccc} \hline
Source & R.A.(B1950) & Dec(B1950) & V$\rm _{LSR}$ \\
       &             &            & (km s$^{-1}$)  \\
\hline
Orion~KL & 05$^h$ 32$^m$ 47$^s$.00 & $-$05\degrees\ 24\amin 30\asec.0 &
  8.0 \\
Sgr~B2(N) & 17$^h$ 44$^m$ 10$^s$.20 & $-$28\degrees\ 21\amin 15\asec.0 &
  64.0 \\
W51~e1/e2 & 19$^h$ 21$^m$ 26$^s$.30 & $+$14\degrees\ 24\amin\ 39\asec.0 &
  60.0 \\
\hline
\end{tabular}
\end{table}

 The heterodyne receivers used were the dual-channel B3 SIS mixers in
 the single-sideband mode (SSB) for the 345 GHz submilliter band
 (315--373 GHz). The sideband rejection was $\sim$13 dB (a factor of
 $\sim$20).  The SSB system temperatures in fair (medium) weather
 conditions -- atmospheric opacity $\tau$(225 GHz) between 0.08 and
 0.12 -- were generally $\sim$400 to 550 K. At 329.9 and 363.1 GHz,
 however, T$_{\rm sys}$ as high as $\sim$700 K and $\sim$900 K were
 recorded, due to poor atmospheric transmission and higher receiver
 temperature.  The half-power beamwidth (HPBW) of the telescope is
 $\sim$14\asec and the main-beam efficiency, $\eta_{mb}$, is 0.63.
 Data were obtained in the position-switching mode with offset
 20\amin~west in azimuth. Pointing and focus was checked regularly at
 a 2-hour interval.  The resultant spectra are on the antenna
 temperature scale, T$_{\rm A}^{\ast}$, which has been corrected from
 chopper wheels calibration for atmospheric transmission and losses
 associated with rearward scattering.  Since the spatial extent of
 each emission source is not known, no main beam correction is
 applied.  Further corrections for the forward scattering and
 spillover efficiency ($\eta_{fss}$ = 0.82) convert the source
 antenna temperature, T$_{\rm A}^{\ast}$, to the source brightness
 temperature, T$_{\rm R}^{\ast}$.

 By employing the Dutch Autocorrelation Spectrometer (DAS) backend
 with two subsystems and a bandwidth of 500 MHz for dual-polarity
 operation, we have a spectral resolution of 756 kHz and a channel
 spacing of 625 kHz.  Typically an integration time of 3 to 4 hours
 was achieved.  Velocity shifting of $\pm$4.5 km s$^{-1}$ with respect
 to the nominal LSR velocity, about $\pm$5 MHz with respect to the
 rest frequency in the 345 GHz band, was executed as a common practice
 during observations in order to neutralize the effect of possible
 low-level gain variations in the DAS and to identify potential
 interlopers from the image sideband.  The JCMT data were reduced
 using the S{\sevensize PECX} spectral line reduction package.

 Observations of submillimeter high-frequency transitions yield smaller
 telescope beams, preferentially sample the warmer and denser regions,
 and help to avoid line confusion with interloping emission from cooler
 envelope material along the line-of-sight.  At hot core temperatures of
 100 K or more, searches in the submillimeter regime are further favoured
 by the fact that the higher frequency transitions are expected to be
 intrinsically stronger.

\addtocounter{table}{1}
\xdef\tabnum{\thetable}
 Good agreement between observed line frequencies and those measured in
 the laboratory, for four or more spectral lines, are the minimum
 requirements for claiming identification of a new interstellar molecule.
 The rotational spectrum of pyrimidine has recently been measured over
 the spectral range 3--337 GHz, and the calculated dipole moment of 
 pyrimidine is $\mu_{tot}=\mu_b$ = 2.39 Debye \citep{kis99}. Hence only
 $b$-type transitions are observable.  Our astronomical search was based
 on the best candidate transitions of pyrimidine in spectral regions free
 from known spectral line contamination.  Four bandheads made up of 
 closely spaced high-J transitions (at higher energy levels) plus 2 pairs
 of low-J doublet lines were observed in a total of 6 different spectral
 bands.  The observed transitions are listed in Table~\thetable; 
 transitions with line strengths smaller than 10.0 are not included.
 Column 1 lists the {\it line} number; each {\it line} may include 
 multiple pyrimidine transitions which are blended into one unresolved,
 single spectral-line feature.

\begin{figure*}
\centerline{\includegraphics[angle=0,width=4.5in]{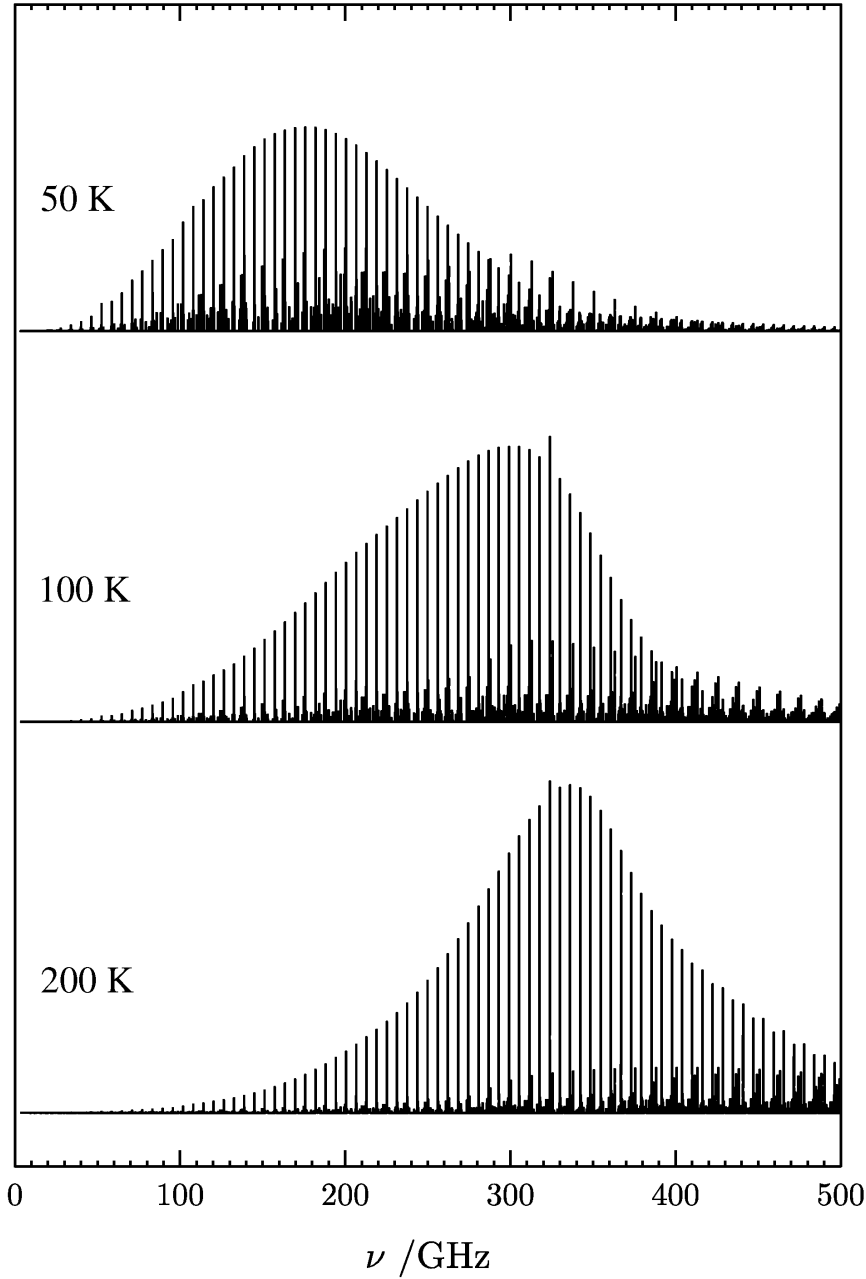}}
\caption{The predicted relative line intensity of the pyrimidine 
 spectra over a frequency range of 0 to 500 GHz at various excitation
 temperatures (50 K, top; 100 K, middle; and 200 K, bottom).  The
 ordinate is in an arbitrary scale.  These spectra are actually contour
 plots with point spacing of 0.5 MHz but have the appearance of stick 
 diagrams because line contours are very narrow relative to the total
 breadth of the plotted spectrum.  Each such stick is thus in fact a 
 very compressed line profile for a spectral blend, and each spectrum
 is in principle 1000000 points long.  The simulation is for a Gaussian
 lineshape with an assumed equivalent linewidth (FWHM) of 7 km s$^{-1}$.
 The {\it apparent} maximum intensity features are at $\sim$180 GHz for
 50 K, $\sim$300 GHz for 100 K and $\sim$335 GHz for 200 K.  The 324
 GHz spectral peak (middle \& bottom) is due to accidental addition
 of some non-band lines to the spectral blend.}\label{fig1}
\end{figure*}

 Figure~\ref{fig1} illustrates the predicted relative line intensity
 in an arbitrary scale of the pyrimidine spectra over a frequency range
 of 0 to 500 GHz at various rotational temperatures (T$_{rot}$ = 50 K,
 the top panel; 100 K, the middle panel; and 200 K, the bottom panel).
 All possible spectral blends are accounted for in the simulation, which
 is made for a Gaussian lineshape with an assumed equivalent linewidth
 (FWHM) of 7 km s$^{-1}$.  Note that the linewidth, when Doppler 
 converted from velocity to frequency, increases with frequency, which
 is conducive to the formation of stronger blends at higher frequencies.

\tablehead{
  \multicolumn{5}{l}{{\bf Table \tabnum.} Pyrimidine transitions observed.} \\
  \hline
  Line  &  Rest Frequency	&  Transition	&
  S$_{\rmn{ul}}^{\star}$	&	E$_{\rmn{l}}^{\dagger}$	   \\
  & (MHz) & J$_{\rmn{K_a,K_c}}$ -- J$_{\rmn{K_a^{\prime},K_c^{\prime}}}$ &
  &  (cm$^{-1}$)	   \\
  \hline
}

\secondtablehead{
  \multicolumn{5}{l}{{\bf Table \tabnum.} (continued).} \\
  \hline
  Line  &  Rest Frequency    	&  Transition	&
  S$_{\rmn{ul}}^{\star}$	&	E$_{\rmn{l}}^{\dagger}$	   \\
  & (MHz) & J$_{\rmn{K_a,K_c}}$ -- J$_{\rmn{K_a^{\prime},K_c^{\prime}}}$ &
  &  (cm$^{-1}$)	   \\
  \hline
}

\tabletail{\hline}

\footnotesize
\begin{supertabular}{ccccc}

1 & 329961.004 &  53$_{1,53}-52_{0,52}$   & 52.48 & 288.87 \nextline
  & 329961.004 &  53$_{0,53}-52_{1,52}$   & 52.48 & 288.87 \nextline
  & 329963.042 &  52$_{2,51}-51_{1,50}$   & 50.45 & 288.46 \nextline
  & 329963.042 &  52$_{1,51}-51_{2,50}$   & 50.45 & 288.46 \nextline
  & 329963.507 &  43$_{11,33}-42_{10,32}$ & 32.73 & 266.24 \nextline
  & 329963.507 &  43$_{10,33}-42_{11,32}$ & 32.73 & 266.24 \nextline
  & 329963.584 &  42$_{12,31}-41_{11,30}$ & 30.83 & 261.71 \nextline
  & 329963.584 &  42$_{11,31}-41_{12,30}$ & 30.83 & 261.71 \nextline
  & 329963.825 &  44$_{10,35}-43_{9,34}$  & 34.65 & 270.36 \nextline
  & 329963.825 &  44$_{9,35}-43_{10,34}$  & 34.65 & 270.36 \nextline
  & 329964.312 &  41$_{12,29}-40_{13,28}$ & 28.95 & 256.77 \nextline
  & 329964.312 &  41$_{13,29}-40_{12,28}$ & 28.95 & 256.77 \nextline
  & 329964.345 &  45$_{9,37}-44_{8,36}$   & 36.58 & 274.06 \nextline
  & 329964.345 &  45$_{8,37}-44_{9,36}$   & 36.58 & 274.06 \nextline
  & 329964.448 &  51$_{3,49}-50_{2,48}$   & 48.43 & 287.64 \nextline
  & 329964.448 &  51$_{2,49}-50_{3,48}$   & 48.43 & 287.64 \nextline
  & 329964.911 &  46$_{8,39}-45_{7,38}$   & 38.52 & 277.36 \nextline
  & 329964.911 &  46$_{7,39}-45_{8,38}$   & 38.52 & 277.36 \nextline
  & 329965.303 &  50$_{3,47}-49_{4,46}$   & 46.43 & 286.41 \nextline
  & 329965.303 &  50$_{4,47}-49_{3,46}$   & 46.43 & 286.41 \nextline
  & 329965.396 &  47$_{7,41}-46_{6,40}$   & 40.48 & 280.24 \nextline
  & 329965.396 &  47$_{6,41}-46_{7,40}$   & 40.48 & 280.24 \nextline
  & 329965.687 &  49$_{4,45}-48_{5,44}$   & 44.43 & 284.76 \nextline
  & 329965.687 &  49$_{5,45}-48_{4,44}$   & 44.43 & 284.76 \nextline
  & 329965.688 &  48$_{6,43}-47_{5,42}$   & 42.45 & 282.71 \nextline
  & 329965.688 &  48$_{5,43}-47_{6,42}$   & 42.45 & 282.71 \nextline
  & 329966.047 &  40$_{14,27}-39_{13,26}$ & 27.08 & 251.42 \nextline
  & 329966.047 &  40$_{13,27}-39_{14,26}$ & 27.08 & 251.42 \nextline
  &	       &			  &	  &	   \nextline
2 & 336125.535 &  54$_{1,54}-53_{0,53}$   & 53.48 & 299.88 \nextline
  & 336125.535 &  54$_{0,54}-53_{1,53}$   & 53.48 & 299.88 \nextline
  & 336127.005 &  43$_{12,32}-42_{11,31}$ & 31.81 & 272.72 \nextline
  & 336127.005 &  43$_{11,32}-42_{12,31}$ & 31.81 & 272.72 \nextline
  & 336127.169 &  44$_{10,34}-43_{11,33}$ & 33.72 & 277.25 \nextline
  & 336127.169 &  44$_{11,34}-43_{10,33}$ & 33.72 & 277.25 \nextline
  & 336127.422 &  42$_{12,30}-41_{13,29}$ & 29.92 & 267.78 \nextline
  & 336127.422 &  42$_{13,30}-41_{12,29}$ & 29.92 & 267.78 \nextline
  & 336127.563 &  53$_{2,52}-52_{1,51}$   & 51.45 & 299.47 \nextline
  & 336127.563 &  53$_{1,52}-52_{2,51}$   & 51.45 & 299.47 \nextline
  & 336127.680 &  45$_{10,36}-44_{9,35}$  & 35.63 & 281.37 \nextline
  & 336127.680 &  45$_{9,36}-44_{10,35}$  & 35.63 & 281.37 \nextline
  & 336128.356 &  46$_{8,38}-45_{9,37}$   & 37.57 & 285.07 \nextline
  & 336128.356 &  46$_{9,38}-45_{8,37}$   & 37.57 & 285.07 \nextline
  & 336128.740 &  41$_{14,28}-40_{13,27}$ & 28.05 & 262.43 \nextline
  & 336128.740 &  41$_{13,28}-40_{14,27}$ & 28.05 & 262.43 \nextline
  & 336128.946 &  52$_{3,50}-51_{2,49}$   & 49.43 & 298.65 \nextline
  & 336128.946 &  52$_{2,50}-51_{3,49}$   & 49.43 & 298.65 \nextline
  & 336129.052 &  47$_{8,40}-46_{7,39}$   & 39.51 & 288.36 \nextline
  & 336129.052 &  47$_{7,40}-46_{8,39}$   & 39.51 & 288.36 \nextline
  & 336129.643 &  48$_{6,42}-47_{7,41}$   & 41.47 & 291.24 \nextline
  & 336129.643 &  48$_{7,42}-47_{6,41}$   & 41.47 & 291.24 \nextline
  & 336129.762 &  51$_{3,48}-50_{4,47}$   & 47.43 & 297.42 \nextline
  & 336129.762 &  51$_{4,48}-50_{3,47}$   & 47.43 & 297.42 \nextline
  & 336130.023 &  49$_{6,44}-48_{5,43}$   & 43.45 & 293.71 \nextline
  & 336130.023 &  49$_{5,44}-48_{6,43}$   & 43.45 & 293.71 \nextline
  & 336130.092 &  50$_{5,46}-49_{4,45}$   & 45.43 & 295.77 \nextline
  & 336130.092 &  50$_{4,46}-49_{5,45}$   & 45.43 & 295.77 \nextline
  & 336131.417 &  40$_{15,26}-39_{14,25}$ & 26.20 & 256.66 \nextline
  & 336131.417 &  40$_{14,26}-39_{15,25}$ & 26.20 & 256.66 \nextline
  &	       &			  &	  &	   \nextline
3 & 338017.612 &  27$_{27,1}-26_{26,0}$   & 24.97 & 146.26 \nextline
  & 338018.346 &  27$_{27,0}-26_{26,1}$   & 24.97 & 146.26 \nextline
  &	       &			  &	  &	   \nextline
4 & 342289.902 &  55$_{1,55}-54_{0,54}$   & 54.48 & 311.09 \nextline
  & 342289.902 &  55$_{0,55}-54_{1,54}$   & 54.48 & 311.09 \nextline
  & 342290.286 &  44$_{12,33}-43_{11,32}$ & 32.80 & 283.93 \nextline
  & 342290.286 &  44$_{11,33}-43_{12,32}$ & 32.80 & 283.93 \nextline
  & 342290.410 &  43$_{12,31}-42_{13,30}$ & 30.91 & 278.99 \nextline
  & 342290.410 &  43$_{13,31}-42_{12,30}$ & 30.91 & 278.99 \nextline
  & 342290.681 &  45$_{10,35}-44_{11,34}$ & 34.70 & 288.46 \nextline
  & 342290.681 &  45$_{11,35}-44_{10,34}$ & 34.70 & 288.46 \nextline
  & 342291.342 &  42$_{13,29}-41_{14,28}$ & 29.03 & 273.64 \nextline
  & 342291.342 &  42$_{14,29}-41_{13,28}$ & 29.03 & 273.64 \nextline
  & 342291.379 &  46$_{9,37}-45_{10,36}$  & 36.62 & 292.58 \nextline
  & 342291.379 &  46$_{10,37}-45_{9,36}$  & 36.62 & 292.58 \nextline
  & 342291.921 &  54$_{1,53}-53_{2,52}$   & 52.45 & 310.68 \nextline
  & 342291.921 &  54$_{2,53}-53_{1,52}$   & 52.45 & 310.68 \nextline
  & 342292.209 &  47$_{8,39}-46_{9,38}$   & 38.56 & 296.28 \nextline
  & 342292.209 &  47$_{9,39}-46_{8,38}$   & 38.56 & 296.28 \nextline
  & 342293.032 &  48$_{8,41}-47_{7,40}$   & 40.51 & 299.58 \nextline
  & 342293.032 &  48$_{7,41}-47_{8,40}$   & 40.51 & 299.58 \nextline
  & 342293.280 &  53$_{3,51}-52_{2,50}$   & 50.43 & 309.86 \nextline
  & 342293.280 &  53$_{2,51}-52_{3,50}$   & 50.43 & 309.86 \nextline
  & 342293.488 &  41$_{14,27}-40_{15,26}$ & 27.17 & 267.87 \nextline
  & 342293.488 &  41$_{15,27}-40_{14,26}$ & 27.17 & 267.87 \nextline
  & 342293.729 &  49$_{6,43}-48_{7,42}$   & 42.47 & 302.46 \nextline
  & 342293.729 &  49$_{7,43}-48_{6,42}$   & 42.47 & 302.46 \nextline
  & 342294.057 &  52$_{3,49}-51_{4,48}$   & 48.43 & 308.63 \nextline
  & 342294.057 &  52$_{4,49}-51_{3,48}$   & 48.43 & 308.63 \nextline
  & 342294.195 &  50$_{5,45}-49_{6,44}$   & 44.44 & 304.93 \nextline
  & 342294.195 &  50$_{6,45}-49_{5,44}$   & 44.44 & 304.93 \nextline
  & 342294.334 &  51$_{5,47}-50_{4,46}$   & 46.43 & 306.98 \nextline
  & 342294.334 &  51$_{4,47}-50_{5,46}$   & 46.43 & 306.98 \nextline
  & 342297.447 &  40$_{15,25}-39_{16,24}$ & 25.33 & 261.70 \nextline
  & 342297.447 &  40$_{16,25}-39_{15,24}$ & 25.33 & 261.70 \nextline
  &	       &			  &	  &	   \nextline
5 & 348453.267 &  44$_{12,32}-43_{13,31}$ & 31.89 & 290.41 \nextline
  & 348453.267 &  44$_{13,32}-43_{12,31}$ & 31.89 & 290.41 \nextline
  & 348453.422 &  45$_{12,34}-44_{11,33}$ & 33.78 & 295.35 \nextline
  & 348453.422 &  45$_{11,34}-44_{12,33}$ & 33.78 & 295.35 \nextline
  & 348453.838 &  43$_{14,30}-42_{13,29}$ & 30.01 & 285.06 \nextline
  & 348453.838 &  43$_{13,30}-42_{14,29}$ & 30.01 & 285.06 \nextline
  & 348454.040 &  46$_{11,36}-45_{10,35}$ & 35.69 & 299.88 \nextline
  & 348454.040 &  46$_{10,36}-45_{11,35}$ & 35.69 & 299.88 \nextline
  & 348454.101 &  56$_{1,56}-55_{0,55}$   & 55.48 & 322.51 \nextline
  & 348454.101 &  56$_{0,56}-55_{1,55}$   & 55.48 & 322.51 \nextline
  & 348454.919 &  47$_{10,38}-46_{9,37}$  & 37.61 & 304.00 \nextline
  & 348454.919 &  47$_{9,38}-46_{10,37}$  & 37.61 & 304.00 \nextline
  & 348455.498 &  42$_{14,28}-41_{15,27}$ & 28.15 & 279.29 \nextline
  & 348455.498 &  42$_{15,28}-41_{14,27}$ & 28.15 & 279.29 \nextline
  & 348455.899 &  48$_{9,40}-47_{8,39}$   & 39.55 & 307.70 \nextline
  & 348455.899 &  48$_{8,40}-47_{9,39}$   & 39.55 & 307.70 \nextline
  & 348456.112 &  55$_{2,54}-54_{1,53}$   & 53.45 & 322.10 \nextline
  & 348456.112 &  55$_{1,54}-54_{2,53}$   & 53.45 & 322.10 \nextline
  & 348456.848 &  49$_{8,42}-48_{7,41}$   & 41.50 & 310.99 \nextline
  & 348456.848 &  49$_{7,42}-48_{8,41}$   & 41.50 & 310.99 \nextline
  & 348457.447 &  54$_{2,52}-53_{3,51}$   & 51.43 & 321.28 \nextline
  & 348457.447 &  54$_{3,52}-53_{2,51}$   & 51.43 & 321.28 \nextline
  & 348457.650 &  50$_{6,44}-49_{7,43}$   & 43.47 & 313.87 \nextline
  & 348457.650 &  50$_{7,44}-49_{6,43}$   & 43.47 & 313.87 \nextline
  & 348458.186 &  53$_{3,50}-52_{4,49}$   & 49.42 & 320.04 \nextline
  & 348458.186 &  53$_{4,50}-52_{3,49}$   & 49.42 & 320.04 \nextline
  & 348458.202 &  51$_{5,46}-50_{6,45}$   & 45.44 & 316.34 \nextline
  & 348458.202 &  51$_{6,46}-50_{5,45}$   & 45.44 & 316.34 \nextline
  & 348458.410 &  52$_{4,48}-51_{5,47}$   & 47.43 & 318.40 \nextline
  & 348458.410 &  52$_{5,48}-51_{4,47}$   & 47.43 & 318.40 \nextline
  & 348458.771 &  41$_{15,26}-40_{16,25}$ & 26.30 & 273.12 \nextline
  & 348458.771 &  41$_{16,26}-40_{15,25}$ & 26.30 & 273.12 \nextline
  &	       &			  &	  &	   \nextline
6 & 363098.770 &  29$_{29,1}-28_{28,0}$   & 26.97 & 169.23 \nextline
  & 363099.057 &  29$_{29,0}-28_{28,1}$   & 26.97 & 169.23 \nextline
  &	       &			  &	  &	   \nextline

\end{supertabular}

$^\star$ Line Strength.

$^\dagger$ Lower energy level. \\

\normalsize

 The strongest lines arise from blends of many transitions in high-J
 bandheads.  These bandheads become more compact and have more lines
 at higher frequencies.
 On the other hand, following Boltzmann population distribution, 
 intensities of individual transitions forming the spectral blends,
 or {\it bands}, maximize at appreciably lower frequencies 
 (henceforth the Boltzmann maxima) than the maxima apparent in the
 spectral profiles in Figure~\ref{fig1}.  At T$_{rot}$ = 100 K, for
 example, the strongest individual band line, i.e., the Boltzmann 
 maximum, though not obvious in the spectrum (the middle panel of
 Figure~\ref{fig1}), is near 195 GHz.  Nevertheless, because at
 such frequencies the band compression effect becomes dominant and
 compensates for the intensity decrease of transitions at frequencies
 beyond the Boltzmann maximum, the overall band profile reaches an
 intensity maximum at the much higher frequency of $\sim$300 GHz.
 Similarly at T$_{rot}$ = 50 K and 200K, the Boltzmann maxima are
 at $\sim$135 GHz and $\sim$285 GHz, respectively, while at higher
 frequencies where band compression dominates, the {\it apparent} 
 maximum intensity features for a linewidth of 7 km s$^{-1}$ are
 predicted to be at $\sim$180 GHz and $\sim$335 GHz, accordingly.
 The 324 GHz spectral blend, which protrudes above the general 
 profile and is seen in the middle and bottom panels of 
 Figure~\ref{fig1}, arises from accidental addition of some 
 non-band lines to the spectral blend.  The prevailing spectral
 features at 100 K and 200 K are compact high-J R-type bandheads of
 the type observed in our search (see Table 2).  For an excitation
 temperature near 100 K, it is clear that searches in the spectral
 region near 1-mm (300 GHz) should provide the best chance for a 
 detection.

\section{Results}

 Table~\ref{tbl-3} lists the important measured physical parameters of
 all the pyrimidine lines observed.  The pyrimidine lines are
 separated into three sources according to the actual observations
 (Column 1).  Column 2 gives the line numbers, which correspond to the
 line numbers shown in Column 1 of Table~\thetable.  Most of the 
 pyrimidine lines are only observed in one or two sources but not in all.  
 The fact that most of the data are fragmentary is mainly due to 
 scheduling difficulties at the JCMT, leading to limited or uncertain
 availability of observing time for different sources.

 In total, 6 different pyrimidine lines were observed toward the 3 
 target sources.  The peak antenna temperature of the spectral line 
 observed, i.e. the T$_{\rm A}^{\ast}$ upper limit for a nondection,
 is given in Column 3 in mK; ``rms'' indicates the 1-$\sigma$ noise
 level per channel of the spectrum, where 
 $ \sigma \equiv {\rm T}_{\rm A}^{\ast}(rms) $.
 Column 4 lists the upper limits of the total column density N$_{tot}$
 derived.  T$_{\rm A}^{\ast}$, hence also N$_{tot}$, are not listed if
 the target pyrimidine line is completely overwhelmed by nearby {\it
 strong} interlopers.  Self-explanatory comments for each spectral
 line observed in each source are given, when necessary, in Column 5.
 Interlopers, which are largely blended with (denoted as ``(B)'') and
 smeared out the target pyrimidine lines or partially blended (denoted
 as ``(PB)'') with the candidate lines, are also included in Column 5
 in the format {\it molecule/frequency(MHz)}.

 Assuming that the pyrimidine lines are optically thin, in LTE, and
 that the rotational excitation temperature, T$_{rot}$, is much higher
 than the background brightness temperature, the beam-averaged total
 column density of $c$-C$_4$H$_4$N$_2$ can be written as:
\begin{equation}
 N_{tot} ({\rm cm}^{-2}) = (1.669 \times 10^{17}) \frac{W_K {\rm Q}_{rot}}
 {\nu} \left\{ \sum \left[ \frac{{\rm S}_{ul}~\mu_b^2}{{\rm exp} \left(
 \frac{{\rm E}_u}{{\rm T}_{rot}} \right) } \right] \right\} ^{-1}, \label{eqn}
\end{equation}
 where
\begin{equation}
 W_K = \int{{\rm T_R}^\ast~dv} = \int{ \frac {{\rm T_A}^\ast}{\eta_{fss}}~dv}
\end{equation}
 in (K km s$^{-1}$) is the integrated intensity of the spectral {\it
 line} either from a single pyrimidine transition, or from unresolved
 multi-transitions such as in a bandhead.  The rest frequency (MHz) is
 $\nu$, $\mu_b$ the permanent $b$-dipole moment in Debye, S$_{ul}$ the
 line strength, E$_u$ the upper energy level in K and Q$_{rot}$ the
 rotational partition function.  All pyrimidine lines reported here
 contain more than one pyrimidine transition, a summation of
 $ ( {\rm S}_{ul} \mu_b^2 ) / [ {\rm exp} ( {\rm E}_u / {\rm T}_{rot} ) ] $
 over all relevant transitions was thus applied.  In addition, a
 rotational temperature T$_{rot}$ = 100 K was assumed for all
 transitions.  In the case of a non-detection, for an
 interloper-blended pyrimidine line, T$_{\rm R}^{\ast}$ is obtained
 from the T$_{\rm A}^{\ast}$ upper limit listed in Column 3 of
 Table~\ref{tbl-3} corrected for $\eta_{fss}$.  If the equivalent
 linewidth could not be unambiguously defined from the observed
 spectrum, the column density limits were evaluated using a 
 representative value of the equivalent linewidth; a value of $\Delta v
 \simeq 10~{\rm km}~{\rm s}^{-1}$ was adopted for all three target
 sources.

\begin{table*}
\centering
\begin{minipage}{180mm}
\caption{The measured physical parameters of the pyrimidine
    lines observed.}\label{tbl-3}
\begin{tabular}{lcccl} \hline
	Source	 	& Line	& T$_{\rm A}^{\ast}$ $\pm$ rms &
	N$_{tot}$	& Comment	\\
		 	&		&   (mK)               &
  (10$^{14}$ cm$^{-2}$)	&		\\
\hline
Sgr B2(N)  &  1 &  ------        & ------  & (B) $ \rm C_2H_5OH$/329954.9;
                                             (B) $ \rm C_2H_5OH$/329956.5;
                                             (B) $ \rm C_2H_5OH$/329968.2 \\
Sgr B2(N)  &  3 &  52.6$\pm$13.5 & $<$1.68 & (B) $ \rm CH_3OCH_3$/338025.5 \\
Sgr B2(N)  &  5 &  ------        & ------  & (B) $ \rm CH_3CHO$/348449.6;
                                             (B) $ \rm C_2H_5CN$/348473 \\
	   &    &		 &	   &				\\
Orion KL   &  2 & 215.7$\pm$13.8 & $<$1.18 & (B) $c$-C$_3{\rm H}_2$/336128.5;
                                             (B) $ \rm C_2H_3CN$/336137.8; \\
           &    &                &         & (PB) $ \rm HCOOCH_3$/336111.3;
                                             (PB) $ \rm SO_2$/336113.5 \\
Orion KL   &  5 & 400.0$\pm$12.6 & $<$3.66 & contaminated by a spectral
                                             feature from the image sideband \\
Orion KL   &  6 &  ------        & ------  & (B) $ \rm CH_3C_3N$/363097.1;
                                             (B) $ \rm Si^{18}O$/363100.7;
                                             (B) $ \rm C_2H_5CN$/363107 \\
	   &    &		 &	   &				\\
W51 e1/e2  &  3 & 140.6$\pm$12.9 & $<$3.14 & (B) $ \rm CH_3OCH_3$/338025.5 \\
W51 e1/e2  &  4 & 631.3$\pm$23.6 & $<$4.79 & detected? (U-342290);
                                             (PB) $ \rm C_2H_3CN$/342286.8 \\
W51 e1/e2  &  5 & 127.5$\pm$12.3 & $<$2.20 & (B) $ \rm CH_3CHO$/348449.6 \\
\hline
\end{tabular}
\end{minipage}
\end{table*}

\begin{figure}
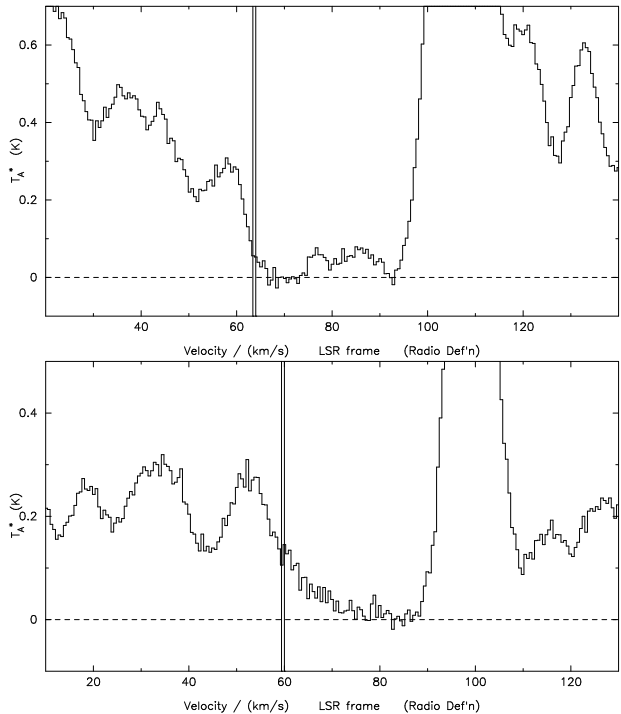

\centerline{\includegraphics[angle=-90,width=3.2in]{fig2a.eps}}
\centerline{\includegraphics[angle=-90,width=3.2in]{fig2b.eps}}
\caption{Sample pyrimidine spectra at 338.0 GHz (Line 3;
 27$_{27,1}$-26$_{26,0}$ and 27$_{27,0}$-26$_{26,1}$) of Sgr B2(N) (top)
 and of W51 e1/e2 (bottom). The two vertical lines mark the two expected
 transitions of pyrimidine. The abscissae give the LSR velocities with 
 respect to the rest frequencies adopted for the observations at the 
 nominal LSR velocities 64.0 km s$^{-1}$ for Sgr B2(N) and 60.0 
 km s$^{-1}$ for W51 e1/e2.}\label{fig2}
\end{figure}

\begin{figure}
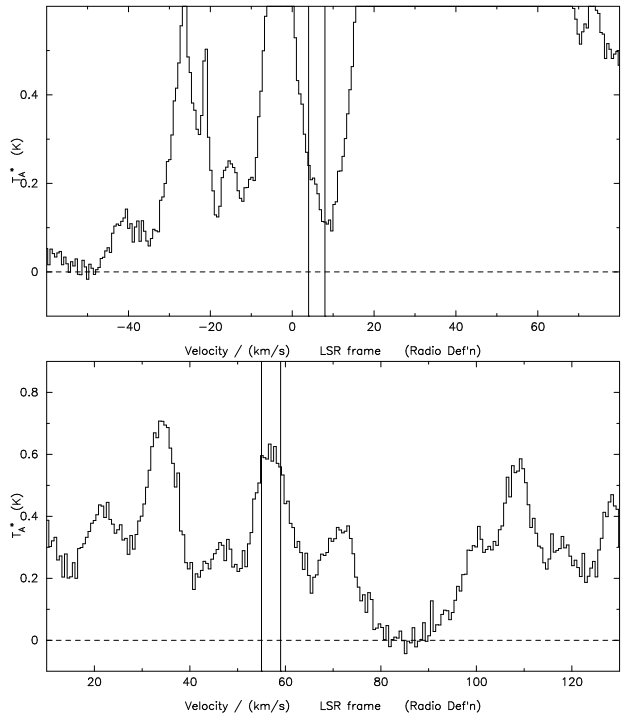

\centerline{\includegraphics[angle=-90,width=3.2in]{fig3a.eps}}
\centerline{\includegraphics[angle=-90,width=3.2in]{fig3b.eps}}
\caption{Sample pyrimidine spectra at the 336.1 GHz band head (Line 2;
 with J $\leq$ 54) of Orion KL (top), and at the 342.3 GHz band head
 (Line 4; with J $\leq$ 55) of W51 e1/e2 (bottom).  The two verticals
 enclose the 336.1 GHz band head (top) and the 342 GHz band head (bottom).
 A {\it tentative} detection of pyrimidine in W51 e1/e2 is visible in
 the bottom panel. The assumed LSR velocities for the rest frequencies
 adopted for the observations are 8.0 and 60.0 km s$^{-1}$ for Orion KL
 and W51 e1/e2, respectively.}\label{fig3}
\end{figure}

 Sample pyrimidine spectra are shown in Figures~\ref{fig2} and
 \ref{fig3}.  With a limited number of spectral lines observed in each
 source, our submillimeter-wave search for pyrimidine did not yield a
 definite detection of pyrimidine in either Sgr B2(N) or Orion KL.  By
 averaging over all observed lines measurable in each source (see
 Table~\ref{tbl-3}), the inferred upper limits on the total column
 density in Sgr B2 and Orion are 1.7$\times$10$^{14}$ cm$^{-2}$ and
 2.4$\times$10$^{14}$ cm$^{-2}$, respectively.

 In the case of W51~e1/e2, there is a spectral feature at $\sim$ 57 km
 s$^{-1}$ with respect to the rest frequency adopted for the
 observation (see the spectrum shown in the lower panel of
 Figure~\ref{fig3}).  This feature coincides precisely with the J $\le$
 55 band-head of pyrimidine between 342289.9 and 342297.4 MHz (Line 4),
 containing 32 transitions at V$_{LSR}$ = 59.0 km s$^{-1}$.  An
 unidentified U-line at 342290.0 MHz was previously reported in a line
 survey of Orion KL \citep{sch97}; unfortunately we did not observe
 either Orion or Sgr~B2 at this particular frequency.

 The total pyrimidine column density in W51 e1/e2 derived solely from
 Line 4 is 4.8$\times$10$^{14}$ cm$^{-2}$; the averaged upper limit to
 the column density from all 3 lines available in W51 is N$_{tot}$
 $\leq$ 3.4$\times$10$^{14}$ cm$^{-2}$. It is interesting to note that
 the pyrimidine column density computed only from Line 4 (assuming
 T$_{rot}$ = 100 K) is just slightly higher (within a factor of 2) than
 the upper limits deduced from the two other, interloper-contaminated,
 lines.  It appears that if Line 4 of W51 is truly a detection, we may
 have expected to detect another bandhead, Line 5, whose transitions
 are at higher frequencies.  It could be argued that the fact that we
 failed to detect Line 5 makes the identification of Line 4 less likely.
 However, Line 5 is expected to be much weaker than Line 4 at an
 excitation temperature lower than 100 K, and so, in the absence of
 accurately measured pyrimidine rotational temperatures for these
 sources, we cannot entirely rule out a true detection in W51.

 Finally, the submillimeter transitions targeted in this search
 possess rather high excitation levels, and so any observable
 pyrimidine emission would mostly originate from regions of high
 temperature and/or density, if radiative excitation is negligible.
 As a result, these excited pyrimidine molecules would have been located
 in regions fairly close to protostellar hot cores within molecular
 clouds.  It is therefore probable that the real source size is
 smaller than the telescope beam ($\sim$14\asec); it is more likely to
 be the case particularly for distant sources such as Sgr B2(N) and
 W51 e1/e2.  Hence, one should note that our observations would suffer
 from beam dilution if the source extent is indeed smaller than the
 beam.  The derived upper limits on the column densities, based on
 the assumption that the source fills the beam, could consequently be
 underestimated.

 The upper limits of fractional abundance of pyrimidine with respect
 to molecular hydrogen,  X(${c-{\rm C_4H_4N_2}}$) =
 N$_{tot}(c-{\rm C_4H_4N_2})/{\rm N_{tot}(H_2)}$, may also be deduced.
 Molecular hydrogen column densities inferred from single-dish
 observations with beam sizes similar to the JCMT are employed, in order
 to determine the beam-averaged abundances more accurately. The H$_2$
 column densities adopted are: $\simeq$ 5 $\times$ 10$^{24}$
 cm$^{-2}$~for Sgr B2(N) \citep{num00}, 8 $\times$ 10$^{23}$ cm$^{-2}$~for
 Orion KL \citep{sut95}, and 1 $\times$ 10$^{24}$ cm$^{-2}$~for W51 e1/e2
 \citep{jaf84}. The upper limits of pyrimidine fractional abundances
 thus estimated are X(${c-{\rm C_4H_4N_2}}$) $\leq$ 3.4 $\times$
 10$^{-11}$ for Sgr B2, 3.0 $\times$ 10$^{-10}$ for Orion, and 3.4
 $\times$ 10$^{-10}$ for W51.

\section{Discussion}

 Our submillimeter search for the nucleic acid building-block
 pyrimidine has not been successful. We have perhaps detected
 one-{\it line} in a single source but this needs to be confirmed. The
 negative result therefore cannot be considered definitive and searches
 at lower frequencies may yet detect pyrimidine in hot cores.  The low
 inferred pyrimidine abundance limits may simply reflect the fact that
 pyrimidine is generally of low abundance in massive star-forming
 cores, and hence it is difficult to populate the high-J energy levels
 (J $\leq$ 56 in our study) which are observable at submillimeter
 wavelengths.  On the other hand, it could also imply that either
 pyrimidine is not evaporated efficiently from dust grains at hot core
 temperatures, or is destroyed easily in the warmest regions in hot
 cores sampled by submillimeter observations.

 It would appear that the best chance for an astronomical detection of
 pyrimidine, and other Nitrogen Heterocycles, is probably in the
 circumstellar envelopes (CSEs) of carbon stars.  AGB and post-AGB
 stars (e.g. IRC+10216, CRL 618 and CRL 2688) are copious producers of
 carbonaceous dust particles.  Recent ISO observations of well-known
 protoplanetary nebulae (PPNe) have uncovered more new organic
 molecules, including the first detection of benzene in CRL 618
 (\citealt{cer01a}, \citeyear{cer01b}).  The initial stages of dust
 formation involve polymerization of acetylene to form benzene and
 subsequent $\rm C_2H_2$ additions lead to large polycyclic aromatic
 hydrocarbon (PAH) molecules (\eg \citealt{che92}).  During this
 reaction sequence other triply-bonded molecules can also add to the
 growing ring structures. In particular, recent theoretical work
 indicates that N atoms can become incorporated in ring structures
 through additions involving HCN \citep{ric01}. However, the kinetics
 of ring growth suggests that single rings containing two N atoms will
 be less favoured than single rings with one N atom (i.e. pyridine).
 Even two-ring compounds containing a single N atom, such as quinoline
 and isoquinoline, the N-substituted analogues of naphthalene, could
 be more abundant than pyrimidine \citep{kis03} and may explain why our
 search appears to have been unsuccessful.  Searches for pyridine, 
 quinoline and isoquinoline in the molecular envelopes of evolved stars
 are currently underway.

 As an alternative to ring formation in neutral-neutral reactions, 
 \cite{woo02} have shown that ion-molecule reactions are able to account
 for the abundance of benzene observed in CRL 618.  The growth of larger
 rings, and incorporation of heteroatoms into them, has not yet been
 considered in circumstellar ion-molecule chemistry.  \cite{woo03}
 have proposed one neutral process, involving benzene and CN, that could
 lead to nitrogen being added to ring structures.  However, the product 
 of this reaction has the nitrogen present in a side-group (i.e. 
 benzonitrile, $c$-$\rm C_6H_5CN$) and not bonded into the ring.

 Another possible reason for a low interstellar pyrimidine abundance
 concerns its photostability.  The infrared spectra and photostability
 of pyrimidine have recently been measured in an Ar matrix at 10 K 
 \citep{pee03}.  The stability of pyrimidine (which does not have real
 aromatic properties) against UV photolysis is rather limited and far
 below that of mono and polycyclic aromatic rings. These experiments 
 appear to indicate that any circumstellar pyrimidine would be easily
 destroyed by UV photons, or cosmic ray particles, soon after delivery
 to the interstellar medium.  In this case, any interstellar pyrimidine
 would have to be produced in dark molecular clouds.

\section{Conclusion}

 To strengthen the role of prebiotic interstellar matter in
 Astrobiology, we have searched for 6 pyrimidine lines in three
 massive star-forming regions: Sgr B2(N), Orion KL and W51 e1/e2.  Our
 search was unsuccessful and did not yield a conclusive result, with
 only one potential single-{\it line} detection. The abundance limits
 inferred are $\leq$ 3.4 $\times$ 10$^{-11}$ for Sgr B2, 3.0 $\times$
 10$^{-10}$ for Orion, and 3.4 $\times$ 10$^{-10}$ for W51. Our
 negative result may simply reflect the fact that interstellar
 pyrimidine is of low abundance.  Sources that are ongoing sites of
 carbon dust formation, such as the C-rich envelopes of AGB and
 post-AGB stars, probably present the best opportunity for detecting
 pyrimidine and other nitrogen heterocycles.

\section{Acknowledgements}

 We would like to thank the referee, Tom J. Millar, for his useful
 comments and suggestions.
 The research of YJK was supported by NSC grants 90-2112-M-003-012 and
 91-2112-M-003-016. This work was supported by NASA's Exobiology Program,
 through NASA Ames Interchange NCC2-1162, and by the Netherlands Research
 School for Astronomy (NOVA).  We wish to thank Remo P.J. Tilanus for his
 kind support and help while observing at the JCMT.

\end{document}